\documentclass[numreferences]{kluwer}

\usepackage{amssymb}
\usepackage{klucite}
\usepackage{graphics}
\usepackage{graphicx}

\newcommand{\greeksym}[1]{{\usefont{U}{psy}{m}{n}#1}}
\newcommand{\umu}{\mbox{\greeksym{m}}}

\newdisplay{guess}{Conjecture}

\begin{document}
\begin{article}
\begin{opening}
\title{Nanofriction mechanisms derived from the dependence of friction on load and sliding velocity from air to UHV on hydrophilic silicon}
\author{Andreas \surname{Opitz}
\thanks{Corresponding author - email: andreas.opitz@physik.uni-ausgburg.de}
\thanks{Present Address: Institut f\"ur Physik, Universit\"at Augsburg, 86153 Augsburg, Germany}}
\institute{Institut f\"ur Physik, Technische Universit\"at Ilmenau, PF 100565, 98684 Ilmenau, Germany} %%
\author{Syed Imad-Uddin \surname{Ahmed}}
\institute{CSEM Centre Suisse d'Electronique et de Microtechnique SA, Rue Jaquet Droz 1, CH-2007 Neuch$\rm{\hat a}$tel, Switzerland} %%
\author{Matthias \surname{Scherge}}
\institute{IAVF Antriebstechnik AG, Im Schlehert 32, 76187 Karlsruhe, Germany} %%
\author{Juergen A. \surname{Schaefer}}
\institute{Institut f\"ur Physik and Zentrum f\"ur Mikro- und Nanotechnologien, Technische Universit\"at Ilmenau, PF 100565, 98684 Ilmenau, Germany} %%

\runningauthor{A. Opitz et al.} %%
\runningtitle{Nanofriction mechanisms derived from the dependence on load and sliding velocity} %%

\date{\today}

\begin{abstract}
This paper examines friction as a function of the sliding velocity
and applied normal load from air to UHV in a scanning force
microscope (SFM) experiment in which a sharp silicon tip slides
against a flat Si(100) sample. Under ambient conditions, both
surfaces are covered by a native oxide, which is hydrophilic.
During pump-down in the vacuum chamber housing the SFM, the
behavior of friction as a function of the applied normal load and
the sliding velocity undergoes a change. By analyzing these
changes it is possible to identify three distinct friction regimes
with corresponding contact properties: (a) friction dominated by
the additional normal forces induced by capillarity due to the
presence of thick water films, (b) higher drag force from ordering
effects present in thin water layers and (c) low friction due to
direct solid-solid contact for the sample with the counterbody.
Depending on environmental conditions and the applied normal load,
all three mechanisms may be present at one time. Their individual
contributions can be identified by investigating the dependence of
friction on the applied normal load as well as on the sliding
velocity in different pressure regimes, thus providing information
about nanoscale friction mechanisms.
\end{abstract}


\end{opening}

\section{Introduction}

In microelectromechanical systems (MEMS), the motion of contacting
parts is strongly influenced by the tribological properties of the
system. Adhesion and friction play a critical role in determining
the operational reliability and lifetimes of such systems
\cite{mems1}. One of the main causes of adhesion and friction
problems of MEMS is the presence of a thin water film present on
most surfaces. This is particularly true of silicon, the most
popular of all the MEMS materials. Freshly exposed silicon rapidly
oxidizes in air forming a nm thick hydrophilic oxide layer, which
can ultimately dictate the performance of the entire device. Thus,
it is important to thoroughly understand the adhesion and friction
properties of contacting surfaces down to the micro and
nanoscales. This knowledge will facilitate the fabrication of next
generation micro and nanodevices with tailored tribological
characteristics.

Previous investigations have examined different aspects of
adhesion and friction properties of materials on the micro and
nanoscale. The microfrictional properties of a tribosystem
consisting of a 1~mm sapphire sphere sliding against a flat
silicon sample was investigated from air to ultrahigh vacuum (UHV)
\cite{msbook,ms1}. This study showed that friction initially
decreased exponentially as a function of the pump-down time.
Later, as more water desorbed from the tribocontact distinct peaks
with higher friction were seen in the friction versus pump-down
time curve. These friction peaks were characterized by the
occurrence of distinct stick/slip friction processes and regions
between them that were absent from such effects. It was postulated
that the peaks with stick/slip friction was caused by the
solid-like friction of thin ordered water films confined between
the ball and sample. It was further postulated that friction peaks
occur whenever an ordering effect occurs due to the formation of a
complete monolayer, which results in solid-like friction. In the
transition period between complete monolayers, the molecules were
judged to be not ordered. In this regime, friction was
characterized only by small fluctuations and was liquid-like.
Another group determined the coefficient of friction for two flat
silicon samples \cite{deulin} in a simple experiment. The lower
sample was tilted until the upper sample just begins to slide. By
recording the tilt angle, $\alpha$, when sliding occurred the
coefficient of friction, $\mu$ was calculated from the
relationship: $\mu = \tan \alpha$. The experiment showed that,
when water coverage on the surface is reduced due to desorption in
vacuum, friction is first reduced due to decreasing capillarity
and is then followed by increased friction. Similar experimental
results were obtained in experiments on single-crystalline Ni(100)
\cite{gellman} where friction was measured for decreasing
thickness of an ethanol film in UHV. Smooth sliding at a coverage
of about 300~monolayers changed into stick/slip sliding when the
ethanol film was reduced to around a monolayer. This was then
followed by strong sticking in the submonolayer range. The contact
pressures in the above experiments were of the order of tens or at
most hundreds of MPa, which is not extremely high compared to most
macro and even nano contact situations. Due to this
characteristic, a liquid layer is present at all times on the
surfaces between contacting parts and the confined liquid films
are responsible for causing stick/slip effects etc. In contrast,
in a typical SFM contact the contact area is much smaller
($\sim$20~nm$^2$). With a normal force on the order of 50~nN, the
contact pressure comes out to be in the range of several GPa. This
unique contact feature results in a tribological behavior that is
different from the microfrictional studies discussed above.

In a nanotribological study, friction as a function of the sliding
velocity, relative humidity and wettability was examined by Riedo
et al. \cite{riedo}. In this study, a linear dependence of
friction force vs. logarithm of sliding velocity was observed. The
slope for hydrophilic surfaces was determined to be negative and
for higher sliding velocities the friction force decreased; the
slope was reduced at lower humidities. The results were
interpreted to indicate that the influence of capillary force is
reduced with decreasing humidity and at higher sliding velocities.
It was also shown that the slope for hydrophobic surfaces is
positive and does not change with varying humidity.

A nanotribological investigation, in which the function of
friction force and water film thickness vs. residual gas pressure
during pump down in a vacuum chamber, was conducted by Opitz et
al. \cite{opitz-ss}. In this report, different friction and
contact regimes were identified. The water film thickness for
hydrophilic native oxide covered silicon samples, measured by
distance dynamic force spectroscopy, was found to reduce from
about 2.6~nm to 0.7~nm with decreasing residual gas pressure in
three distinct regimes. Further measurements showed the complete
removal of water due to the combined effect of the vacuum
environment and friction-induced desorption. The measured friction
force also showed three regimes. It was seen that the reduction of
the friction force corresponds to decrease of the water film
thickness.

This work extends the knowledge gained in the above mentioned
study \cite{opitz-ss} in that it examines the dependence of
friction on velocity and on an externally applied normal load and
uses this information to derive nanofriction mechanisms and basic
nanocontact properties of the tribosystem. A continually
decreasing water film thickness was produced by inserting the
sample in a vacuum system and pumping it down from ambient
pressure to 10$^{-8}$~mbar. Hydrophilic, native oxide covered
silicon was used as sample and counterbody. Distinct regimes of
friction versus the applied normal load and the sliding velocity
could be identified.

\section{Experimental setup}

\subsection{Sample preparation}

Rectangular pieces of p-doped (1-10~$\Omega$cm, B-doped) native
oxide covered Si(100) were used as samples. These had an SFM
determined RMS (Root Mean Square) roughness of about 0.1~nm
measured over a 0.6$\times$0.6 $\umu$m$^2$ area. Due to the
presence of a native oxide on top of the Si(100) surface, the
samples are OH-terminated \cite{opitz-wear} and hydrophilic
\cite{silica,maboudian}. The hydrophilicity was also separately
confirmed by contact angle measurements (not shown). Prior to
insertion into the UHV chamber, the samples were sequentially
cleaned using ultrasonic assistance in isopropanol and methanol
for five minutes each. Afterwards, the samples were thoroughly
rinsed with bi-distilled water. The oxide film measured on such
samples has a thickness of 1.3~nm as determined by X-ray induced
photoelectron spectroscopy (XPS).

\subsection{Friction force microscope}

A UHV vacuum SFM/STM (Omicron Nanotechnology GmbH) system was
operated in contact SFM mode to measure friction between a sharp
silicon tip and a planar Si(100) sample. Measurements were
performed at various stages during evacuation of the system, from
air to 10$^{-8}$ mbar, using a combination of turbo molecular, ion
and titanium sublimation pumps. Starting from air, upon pump-down
the chamber pressure decreased rapidly between 10$^1$ and
10$^{-3}$ mbar. Due to the fact that at this stage thermodynamic
equilibrium does not exist, no measurements for the velocity and
normal force characteristic could be performed. Friction
measurements using contact SFM were performed in lateral mode
registering the friction hysteresis at different applied normal
loads with sliding velocity of 300~nm/s and at different sliding
velocities with an applied normal load of 60~nN. For each friction
measurement the tip was scanned 300~nm in the forward and backward
direction and the friction was determined by taking one half of
the friction hysteresis curve \cite{schwarz}. To attain a high
lateral force resolution, single beam Si cantilevers with a length
of 440~$\umu$m were chosen instead of triangular cantilevers that
are typically used for topographical measurements. These single
beam cantilevers, with a normal spring constant of 0.11~N/m, and a
manufacture-quoted tip radius that was smaller than 15~nm, were
used as received, i.e. in an oxidized (hydrophilic) state.
Calibration of the normal and lateral forces was achieved by
following the procedure developed by Schwarz et al.
\cite{schwarz}, which consists of using the geometrical dimensions
of the tip and determining the magnitude of tip deflection as a
function of the applied load.

\section{Results}

\subsection{Residual gas pressure dependence}

Figure~\ref{fig:1} shows the friction force measured as a function
of the residual gas pressure from air down to 10$^{-8}$~mbar. In
this case, the silicon sample was cleaned as described above and
then thoroughly rinsed with water before inserting it into the
vacuum system. After this preparation process the hydrophilic
sample is covered by a water with a thickness of 2~nm. Initially,
a high friction force was measured (Fig.~\ref{fig:1}). Afterwards,
during pump-down, friction decreased from range (a) to range (b)
in Fig.~\ref{fig:1} to half the initial value. This is due to
water desorption when the residual gas pressure is decreased.
However, a subsequent reduction in the friction also occurred at
lower pressures. It is suggested that this last step (marked (c)
in Fig.~\ref{fig:1}) to lower friction values occurs due to the
combined effect of the vacuum and friction-induced desorption of
water from about 0.7~nm thickness to almost complete water
removal. The resulting contact at this stage is then a direct
solid-solid contact without the presence of an additional water
film.

\subsection{Sliding velocity dependence}

The friction force vs. sliding velocity dependence, characteristic
for the three ranges shown in Fig.~\ref{fig:1}, is depicted in
Fig.~\ref{fig:2}. The sliding velocity is plotted in logarithmic
scale. In this plot a linear dependence is visible. The equation
used to fit this measurement is given as \cite{riedo}

\begin{equation}
  F_\mathrm{R} = F_\mathrm{O} + F_\mathrm{S} \ln \left( \frac{v}{v_0} \right).
  \label{eq:1}
\end{equation}

Here are $F_\mathrm{O}$ the intercept and $F_\mathrm{S}$ the
slope. The constant $v_0$ was fixed to 1~nm/s to keep the units
correct.

Figure~\ref{fig:3} shows the changing of the slope
$F_\mathrm{S}$ as a function of the residual gas pressure. Like
in Figure~\ref{fig:1}, here also the three distinct regions:
(a), (b) and (c) can be easily discerned. The slope changes
from high negative values for regime (a) to positive values in
regime (b) and ends at comparatively small positive values for
regime (c). The boundaries between these regimes are the same
like in the friction vs. residual gas pressure measurements (see
Fig.~\ref{fig:1}).

\subsection{Normal force dependence}

Bowden and Tabor \cite{bowden-tabor} described the dependence of
friction force on the applied normal force with respect to the
contact area. Later, Maugis \cite{maugis} derived a value for the
contact area from the elastic theory of solids and from rupture
mechanics. In this work a parameter $\lambda$ is defined whose
magnitude indicates which borderline case of the general contact
model should be applied. It is defined as

\begin{equation}
  \lambda = \left( \frac{8 R \gamma^2}{\pi K^2 \delta^3} \right)^\frac{1}{3} %%
  \label{eq:2}
\end{equation}

where $\gamma$ is the surface energy, $\delta$ the interaction
length in the Dugdale interaction model, $R$ the tip radius and
$K$ the reduced elastic modulus. Using the values $R=30$~nm
\cite{franzka}, $\gamma_\mathrm{SiO_2}=20$~mJ$\cdot$m$^{-2}$
\cite{maboudian}, \mbox{$K_\mathrm{SiO_2}=50.1$~GPa} \cite{msbook}
and assuming that the interatomic distance, $\delta$ is 0.5~nm the
calculation for $\lambda$ gives 0.07. When $\lambda$ is smaller
than 0.1 the DMT-model \cite{dmt} for the friction force
$F_\mathrm{R}$ can be used:

\begin{equation}
  F_\mathrm{R} = \pi \tau \left( \frac{R}{K} \right) ^\frac{2}{3} \left( F_\mathrm{L} + 4 \pi \gamma R \right) ^\frac{2}{3}. %%
  \label{eq:3}
\end{equation}

Here $\tau$ is the shear stress in the friction contact,
$\gamma$ the surface energy and $F_\mathrm{Ad}=4\pi\gamma R$ the
adhesion force in the DMT-Model. The normal force in this model
is the sum of applied normal load $F_\mathrm{L}$ and the
adhesion force $F_\mathrm{Ad}$.

The dependence of friction force on the applied normal load is
shown in Figure~\ref{fig:4} for the three distinct friction
regimes (as was the case in previous figures). The intercept on
the load axis as well as the values for the friction forces change
from region (a) to (c). To extract more information the measured
curves were fitted using Equ.~\ref{eq:3} and the surface energy
and the shear stress were then evaluated (see Fig.~\ref{fig:5}).
Again a value of $K=50.1$~GPa was used and $R=30$~nm was taken as
the tip radius. This is larger than the manufacture-quoted tip
radius (less than 15~nm) because our studies indicate that the tip
gets blunter during measurement \cite{Guig00,franzka}. The surface
energy, $\gamma$, and the shear stress, $\tau$, (Fig.~\ref{fig:5})
reduce in two steps during pump-down. The three distinct pressure
regimes, however, are still visible.

\section{Discussion}

The dependence of friction on the sliding velocity and applied
normal load for a silicon tip sliding against a silicon sample
changes considerably from wet at air pressure to very dry in
ultrahigh vacuum. The nanofriction between two hydrophilic
surfaces (Fig.~\ref{fig:1}) is reduced in three regimes. The slope
in the friction force vs. logarithm of sliding velocity curves
(Fig.~\ref{fig:3}) describes the friction behavior of the
nanotribological contact. The negative slope in regime (a) shows
the domination of capillary forces. The friction force is
determined by the number of capillary bridges that form in the
nanocontact between tip and sample \cite{riedo}. The number of
capillary necks and the influence of capillary force are reduced
at higher sliding velocities assuming an activation process for
the time to build the water bridges \cite{riedo}. When the water
film thickness is above 2~nm \cite{opitz-ss}, for tribological
purposes water can be considered as bulk water. During pump-down,
the slope changes to positive values after reduction of the
residual gas pressure (and with that the water partial pressure).
At this stage, bulk water desorbs when the water partial pressure
of ambient is lower then the water vapor pressure in regime (b).
The residual water film remaining on the sample surface, has a
film thickness of about 0.7~nm and correlates to a film composed
of 2 ice-like water bilayers \cite{opitz-ss,sfa-isra}. In this
regime, ordering effects of the ice-like bilayers strongly
influence friction. We speculate that friction arises from the
"pushing aside" of these bilayers by the sliding tip, which has
always some solid-solid contact present due to the high applied
normal load (pressure) in this study \cite{opitz-ss}. These
ordering effects result in the positive slope observed in
Fig.~\ref{fig:2}. The decreasing of the slope in region (c) is
attributed to the complete removal of water due to desorption. A
similar small positive slope was also measured on hydrophobic
\cite{liu,riedo,msbook} as well as on atomically flat
\cite{gnecco} surfaces. Thus, our measurements demonstrate the
transition of friction from a wet, hydrophilic surface to a dry
hydrophilic one. From these results one can deduce that the
characteristics of a dry hydrophilic surface are comparable to
those of hydrophobic surfaces.

The frictional behavior derived from the nanocontact conditions is
defined by the structure of the water film present on the surface.
The water film in regime (a) is bulk water on top and an
electrochemical double layer between this bulk water and silicon
oxide surface \cite{msbook}. A comparison between bulk water in
regime (a) and ordered water in (b) (Figure~\ref{fig:2}) shows
that, when an extrapolation of the friction force vs. sliding
velocity curves in regime (a) and (b) is made to higher
velocities, they cross at a velocity of about 2~mm/s. This result
suggests that at this velocity the influence of capillary force
vanishes and friction is dominated by ordering effects of the thin
water layer. In other words, this means that the slope of the
friction vs. logarithm of sliding velocity should reverse at
velocities in the millimeter per second range for the examined
contact parameters. However, since the highest achievable sliding
velocity in the used device is about 14~$\umu$m/s, experimental
verification of this phenomenon was not possible.

According to the DMT-model \cite{dmt}, the surface energy relates
to the adhesion force as $F_\mathrm{Ad} = 4 \pi \gamma R$.
Following this equation, adhesion increases for higher surface
energies. In our experiments, the surface energy undergoes a
reduction in 3 regimes (Fig.~\ref{fig:5}). The difference of
surface energy between regime (a) and (b) is about
122~mJ$\cdot$m$^{-2}$ and is in the order of magnitude of the
surface tension of bulk water ($\gamma_\mathrm{H_2O}=72$~mJ
$\cdot$m$^{-2}$). This confirms that bulk water desorbs during the
transition from regime (a) to (b). The lowest value of surface
energy is reached at pressures lower then 10$^{-7}$~mbar, where it
has a value of 21~mJ$\cdot$m$^{-2}$. This value is comparable to
measurements determined by adhesion occurring in a micro
cantilever array \cite{maboudian}. Another quantity, the shear
stress, describes the lateral force per unit area for two moving
parts in contact. In our experiments, the shear stress changes in
two steps (Fig.~\ref{fig:5}). The highest shear stress is measured
at high water film thicknesses in regime (a). At this stage, the
effects of solid-solid contact, ordering of the electrochemical
double layer and bulk water, resist the motion of the tip. The
removal of the bulk water layer shows decreasing of shear stress
in regime (b). At this stage, interactions due to the solid-solid
contact and the ordering effects of the two ice-like water
bilayers dictate the frictional properties. The lowest shear force
(visible in regime (c)) is present when only solid-solid contact
occurs.

It should be noted that at the applied normal load in this study,
solid-solid contact also occurs in the regime where ordering
effects of thin water double layers are present. Also, these two
effects occur in the capillary force regime where bulk water is
present on the surface.

\section{Conclusion}

\begin{figure}[b]
\centerline{\includegraphics{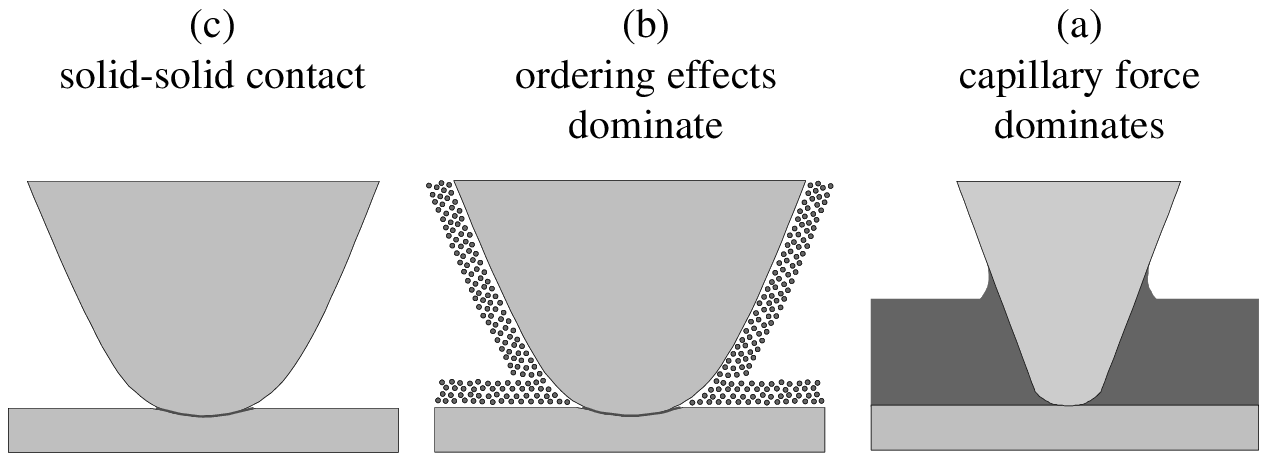}} \caption{} \label{fig:6}
\end{figure}

The friction force was examined as a function of the residual gas
pressure, sliding velocity and applied normal load using contact
SFM. The friction force on hydrophilic silicon undergoes changes
during pump-down in a vacuum system from air to 10$^{-8}$ mbar.
Three distinct friction regimes  with different nanofriction
mechanisms and nanocontact characteristics, were identified. These
regimes are shown schematically in Figure~\ref{fig:6}. First, the
capillary force dominates. Later, during pump-down, most of the
water is removed by vacuum desorption. The remaining water is now
an ordered water double layer and the effects of the ordered water
layer dominate friction. After complete water desorption due to
the combined effects of the vacuum and friction-induced
desorption, only solid-solid contact remains, which exhibits the
lowest friction of all the three friction regimes. In this study,
the water present on the tip and sample surface never acts as a
lubricant.

\begin{acknowledgements}
This work was supported by a grant from the Deutsche
Forschungsgemeinschaft (Project Sche 425/2-4).
\end{acknowledgements}

\clearpage

\begin{center}{\textbf{Figure captions}}\end{center}

\begin{figure}[t]
\centerline{\includegraphics{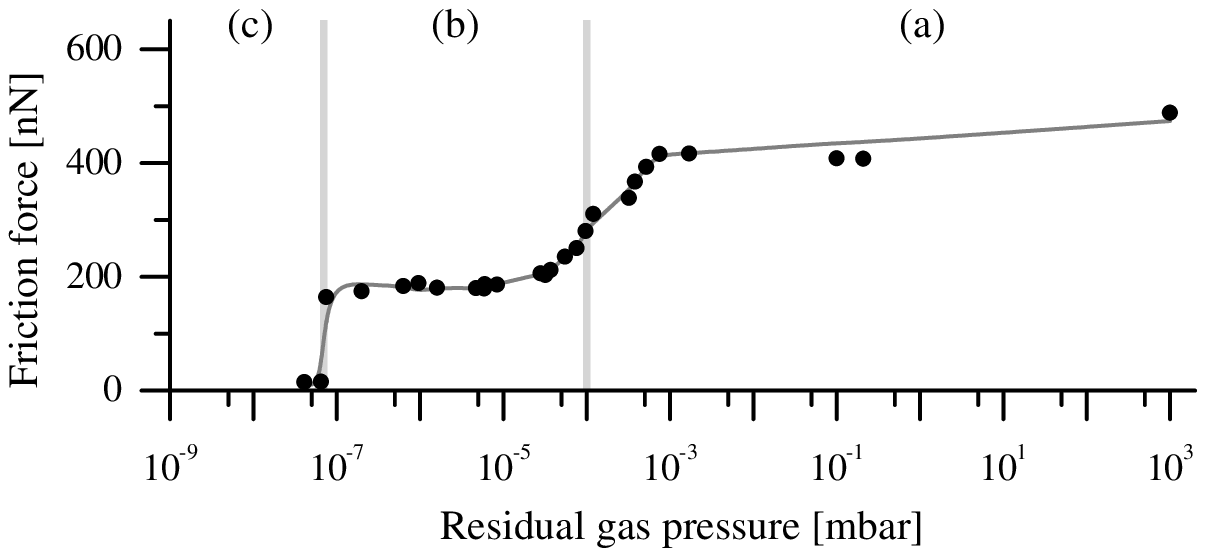}} \caption{} \label{fig:1}
\end{figure}

\textit{Figure~\ref{fig:1}.}

Friction force vs. residual gas pressure during pump down in the
vacuum chamber. The ranges (a), (b) and (c) identify different
friction regimes. The solid line is shown to guide the eye.

\vspace{5mm}

\begin{figure}[b]
\centerline{\includegraphics{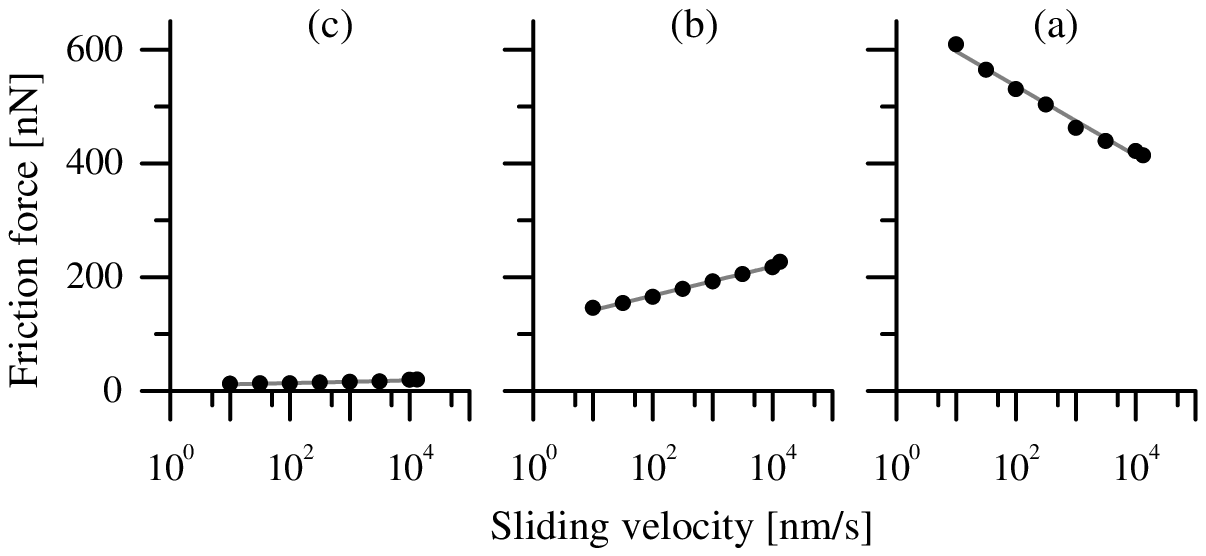}} \caption{} \label{fig:2}
\end{figure}

\textit{Figure~\ref{fig:2}.}

Friction force vs. sliding velocity for the three friction
regimes identified in Fig.~\ref{fig:1}. For every friction
regime a representative measurement is shown. The solid lines
are fits using Equation~\ref{eq:1}.

\vspace{5mm}

\begin{figure}[t]
\centerline{\includegraphics{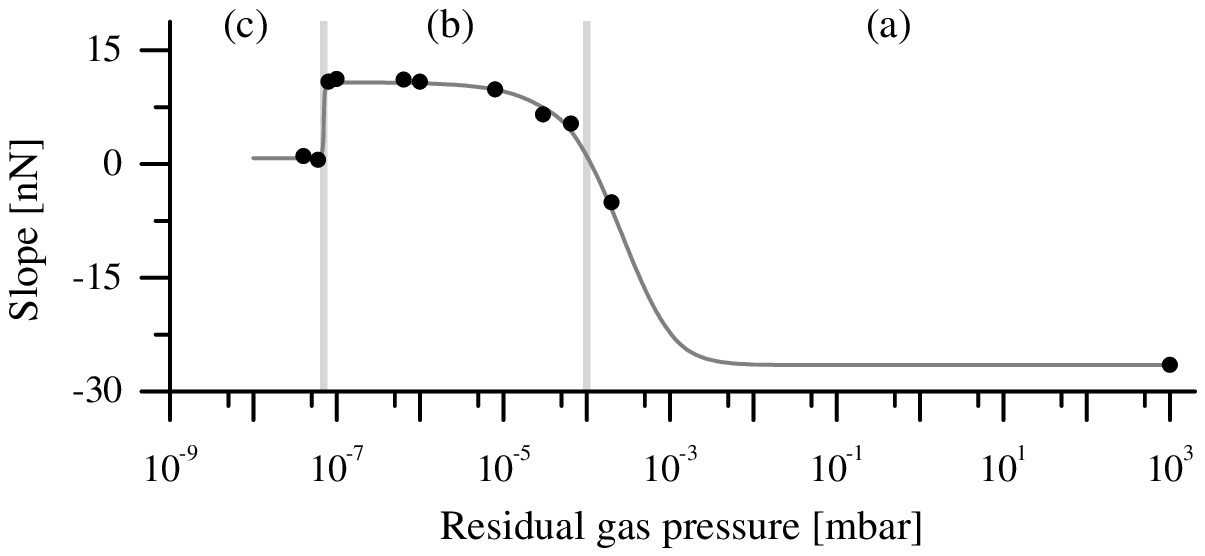}} \caption{} \label{fig:3}
\end{figure}

\textit{Figure~\ref{fig:3}.}

Slope $F_\mathrm{S}$ from Equation~\ref{eq:1} vs. the residual
gas pressure. The three different regimes from Fig.~\ref{fig:1}
are marked using the vertical grey lines. They correspond well
with the change in the slope $F_\mathrm{S}$. The solid line
connecting the data points has been drawn to guide the eye.

\vspace{5mm}

\begin{figure}[b]
\centerline{\includegraphics{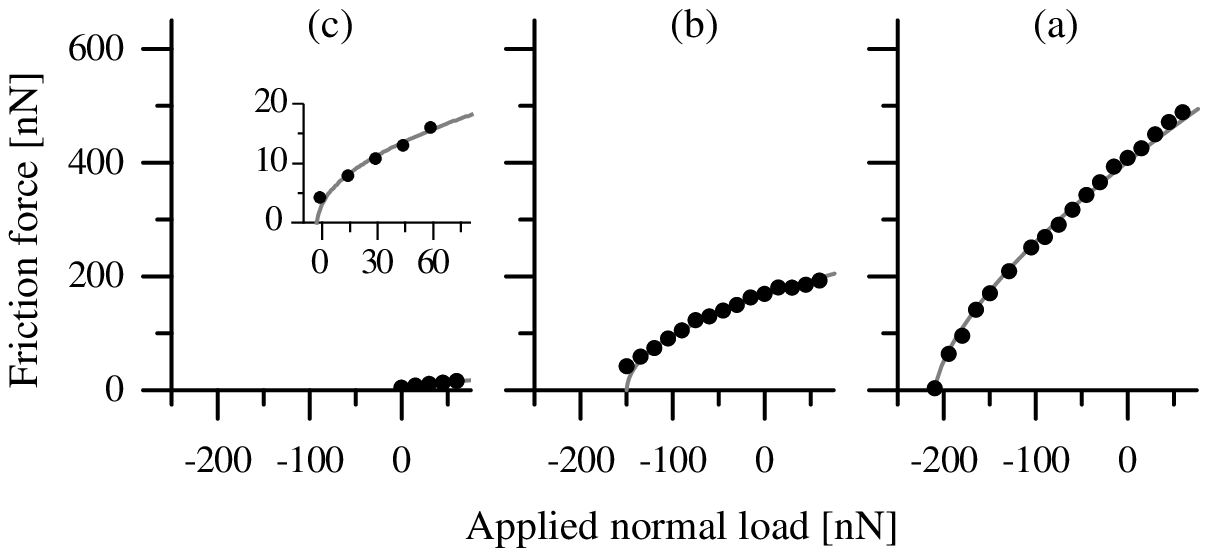}} \caption{} \label{fig:4}
\end{figure}

\textit{Figure~\ref{fig:4}.}

Friction force vs. applied normal load for the three friction
regimes from Fig.~\ref{fig:1}. A representative measurement is
shown for each regime. The solid lines are fits using
Equation~\ref{eq:3}. A zoomed diagram is shown for regime (c) for
better visibility.

\vspace{5mm}

\begin{figure}[t]
\centerline{\includegraphics{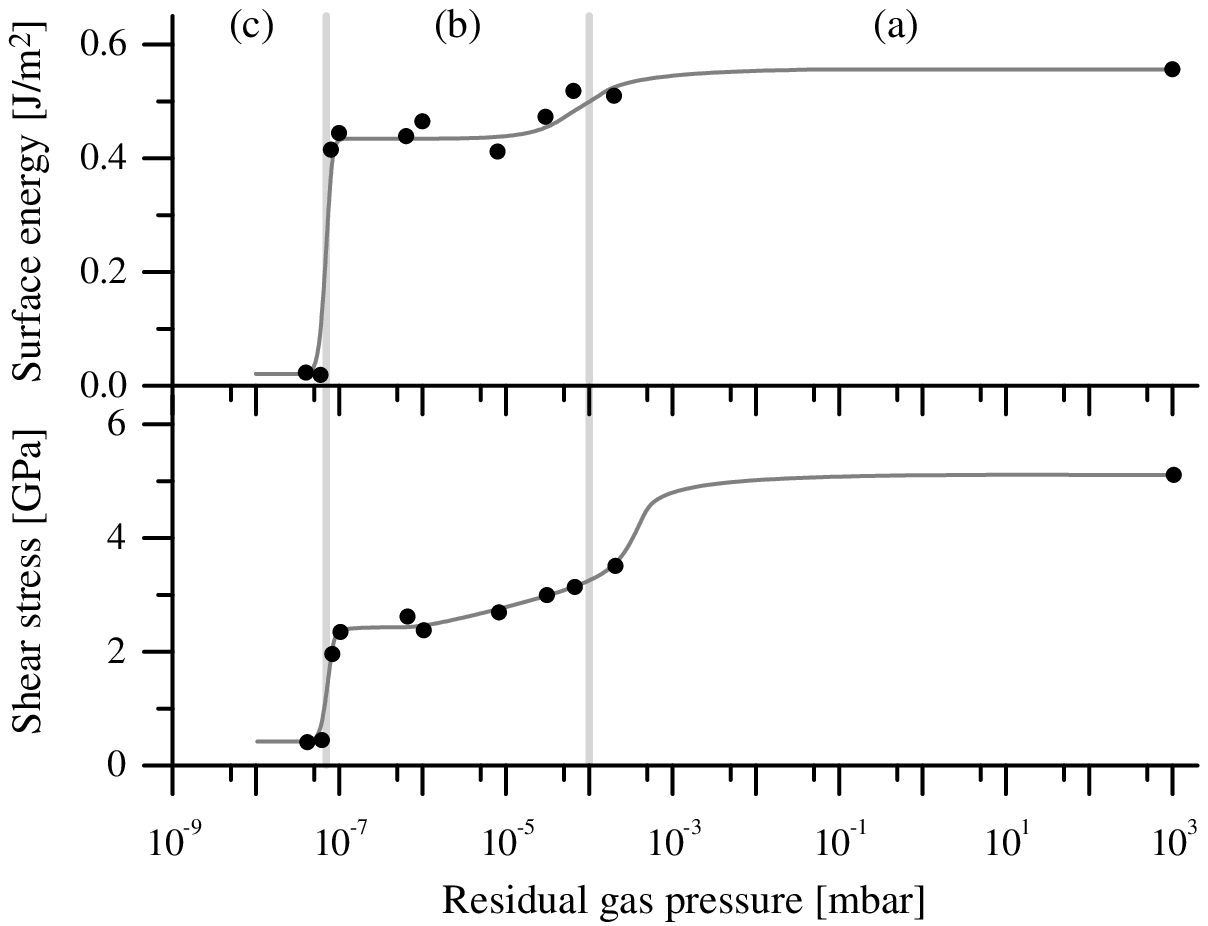}} \caption{} \label{fig:5}
\end{figure}

\textit{Figure~\ref{fig:5}.}

Surface energy and shear stress as determined from
Equation~\ref{eq:3} vs. the residual gas pressure. The three
different friction regimes from Fig.~\ref{fig:1} are marked
using the vertical grey lines. The solid line connecting the
data points has been drawn to guide the eye.

\vspace{5mm}

\textit{Figure~\ref{fig:6}.}

Schematic model of water coverage in various coverage regimes
determined by the friction behavior. While there is only the
solid-solid interaction in regime (c), this interaction along with
ordering effects of water are both present in regime (b) and,
similarly, the solid-solid interaction and ordering effects are
also present in regime (a).

\clearpage

%%\clearpage

%%\clearpage

%%\clearpage

%%\clearpage

%%\clearpage

\end{article}

\begin{thebibliography}{00}

\bibitem{bowden-tabor}
Bowden, F.~P. and D. Tabor: 1950, {\em The friction and
lubrication of solids}.
\newblock Oxford: Clarendon Press.

\bibitem{dmt}
Derjaguin, B.~V., V.~M. Muller, and Y.~P. Toporov: 1975, `Effect
of contact
  deformation on the adhesion particles'.
\newblock {\em J. Colloid Interface Sci.} {\bf 53}, 314.

\bibitem{deulin}
Deulin, E.~A., A.~A. Gatsenko, and B.~A. Loginov: 1999, `Friction
force of
  smooth surfaces of {S}i{O}$_2$--{S}i{O}$_2$ as a function of residual
  pressure'.
\newblock {\em Surf. Sci.} {\bf 433-435}, 288.

\bibitem{franzka}
Franzka, S.: 1999, `Rasterkraftmikroskopische Untersuchungen zur
Reibung von
  Silicium- und Al$_2$O$_3$-Einkristalloberfl\"achen im Kontakt mit Silicium'.
\newblock Dissertation, Universit\"at Karlsruhe.

\bibitem{gellman}
Gellman, A.~J.: 1992, `Lubrication by molecular monolayers at
{N}i-{N}i
  interfaces'.
\newblock {\em J. Vac. Sci. Technol. A} {\bf 10}, 180.

\bibitem{gnecco}
Gnecco, E., R. Bennewitz, T. Gyalog, C. Loppacher, M. Bammerlin,
E. Meyer, and
  H. G\"untherodt: 2000, `Velocity Dependence of Atomic Friction'.
\newblock {\em Phys. Rev. Lett.} {\bf 84}, 1172.

\bibitem{silica}
Iler, R.~K.: 1979, {\em The Chemistry of Silica}.
\newblock Chichester: Wiley Interscience.

\bibitem{sfa-isra}
Israelchvili, J.~N. and R.~M. Pashley: 1983, `Molecular layering
of water at
  surfaces and origin of repulsive hydration forces'.
\newblock {\em Nature} {\bf 306}, 249.

\bibitem{liu}
Liu, H., S.~I.-U. Ahmed, and M. Scherge: 2001, `Microtribological
properties of
  silicon and silicon coated with diamond like carbon, octadecyltrichlorosilan
  and stearic acid cadmium salt films: {A} comparative study'.
\newblock {\em Thin solid films} {\bf 381}, 135.

\bibitem{maboudian}
Maboudian, R.: 1998, `Surface processes in {MEMS} technology'.
\newblock {\em Surf. Sci. Rep.} {\bf 30}, 207.

\bibitem{maugis}
Maugis, D.: 1992, `Adhesion of Spheres: The JKR-DMT Transition
Using a Dugdale
  Model'.
\newblock {\em J. Colloid Interface Sci.} {\bf 150}, 243.

\bibitem{Guig00}
{McGuiggan}, P.~M., J. Zhang, and S.~M. Hsu: 2001, `Comparsion of
friction
  measurements using the atomic force microscope and the surface forces
  apparatus: the issue of scale'.
\newblock {\em Tribol. Lett.} {\bf 10}, 217.

\bibitem{opitz-ss}
Opitz, A., S.~I.-U. Ahmed, J.~A. Schaefer, and M. Scherge: 2002,
`Friction of
  thin water films: a nanotribological study'.
\newblock {\em Surf. Sci.} {\bf 504}, 199.

\bibitem{opitz-wear}
Opitz, A., S.~I.-U. Ahmed, J.~A. Schaefer, and M. Scherge: 2003,
`Nanofriction
  of silicon oxide surfaces covered with thin water films'.
\newblock {\em WEAR} {\bf 254}, 924.

\bibitem{riedo}
Riedo, E., F. L\`{e}vy, and H. Brune: 2002, `Kinetics of Capillary
Condensation
  in Nanoscopic Sliding Friction'.
\newblock {\em Phys. Rev. Lett.} {\bf 88}, 185505.

\bibitem{msbook}
Scherge, M. and S.~N. Gorb: 2001, {\em Biological Micro- and
Nanotribology},
  NanoScience and Technology.
\newblock Berlin: Springer-Verlag.

\bibitem{ms1}
Scherge, M., X. Li, and J.~A. Schaefer: 1999, `The effect of water
on friction
  of {MEMS}'.
\newblock {\em Tribol. Lett.} {\bf 6}, 215.

\bibitem{schwarz}
Schwarz, U.~D., P. K\"oster, and R. Wiesendanger: 1996,
`Quantitative analysis
  of lateral force microscopy experiments'.
\newblock {\em Rev. Sci. Instrum.} {\bf 67}, 2560.

\bibitem{mems1}
Tanner, D.~M., J.~A. Walraven, L.~W. Irwin, M.~T. Dugger, N.~F.
Smith, W.~P.
  Eaton, W.~M. Miller, and S.~L. Miller: 1999, `The Effect of Humidity on the
  Reliability of a Surface Micromachined Microengine'.
\newblock {\em IEEE Int. Rel. Phys. Symp. Proc.} p. 189.

\end{thebibliography}
\end{document}